\documentclass[review]{elsarticle}

\usepackage{lineno}
\modulolinenumbers[5]

\usepackage{graphicx,color}
\usepackage{dcolumn}
\usepackage{amsmath}
\usepackage{amssymb}
\usepackage{epstopdf}
\usepackage{hyperref}

\hypersetup{colorlinks=true,linkcolor=blue,citecolor=blue,urlcolor=blue}

\journal{Physica B}

\newcommand{\slrrtext}  {spin-lattice-relaxation rate}
\newcommand{\urusi}     {URu$_2$Si$_2$}
\newcommand{\slrr}      {$T_1^{-1}$}

\begin{document}

\begin{frontmatter}

\title{$^{29}$Si nuclear magnetic resonance study of URu$_2$Si$_2$ under pressure}

\author[mymainaddress]{K. R. Shirer\corref{mycorrespondingauthor}}
\cortext[mycorrespondingauthor]{Corresponding author}
\ead{krshirer@ucdavis.edu}

\author[mymainaddress]{A. P. Dioguardi}
\author[mymainaddress]{B. T. Bush}
\author[mymainaddress]{J. Crocker}
\author[mymainaddress]{C. H. Lin}
\author[mymainaddress]{P. Klavins}
\author[LANLaddress]{J. C. Cooley}
\author[UCSDaddress]{M. B.  Maple}
\author[NUaddress]{K. B. Chang}
\author[NUaddress]{K. R. Poeppelmeier}
\author[mymainaddress]{N. J. Curro}




\address[mymainaddress]{Department of Physics, University of California, Davis, CA 95616, USA}
\address[LANLaddress]{Los Alamos National Laboratory, Los Alamos, New Mexico 87545, USA}
\address[UCSDaddress]{Department of Physics and Institute for Pure and Applied Physical Sciences, University of California, San Diego, La Jolla, California 92093-0319, USA}
\address[NUaddress]{Northwestern University, 2145 Sheridan Road, Evanston, IL 60208, USA}

\begin{abstract}

We report $^{29}$Si nuclear magnetic resonance measurements of single crystals and aligned powders of URu$_2$Si$_2$ under pressure in the hidden order and paramagnetic phases.  We find that the Knight shift decreases with applied pressure, consistent with previous measurements of the static magnetic susceptibility. Previous measurements of the spin lattice relaxation time revealed a partial suppression of the density of states below 30 K.   This suppression persists under pressure, and the onset temperature is mildly enhanced.

\end{abstract}


\end{frontmatter}


\section{Introduction}

\urusi is a heavy fermion material that has captured the interest of the condensed matter community for over 25 years \cite{PalstraURSdiscovery,MydoshReview}. In addition to an unconventional superconducting state below 1.5K, \urusi\ undergoes a phase transition at $T_{HO} = 17.5K$ \cite{PalstraURSdiscovery,Kasahara2007}, the order parameter of which is yet undetermined. The nature of the hidden order phase remains controversial, but it clearly does not involve magnetic ordering of dipole moments \cite{BroholmURS}.
Extensive neutron scattering work has suggested that it has an itinerant nature and involves some type of Fermi surface instability \cite{BroholmURS,WiebeURS,FlouquetJPSL2010}, and recent Raman spectroscopy experiments indicate that the hidden order consists of a chirality density wave \cite{Kung2015}. It is likely that this exotic phase emerges in this material due to the unusual symmetry of the  crystalline electric field states \cite{HauleURSnature2009}.

An analysis of previous thermodynamic and neutron scattering measurements in the context of hidden order gap fluctuations uncovered the possible existence of a pseudogap occurring at a temperature scale $T_0> T_{HO}$, and recent nuclear magnetic resonance (NMR) experiments at ambient pressure indicated $T_0\sim30K$ \cite{BalatskyURSPG,ShirerURSPRB2012}. The presence of anomalies between 25-30K in magnetic susceptibility \cite{MapleURu2Si2}, point contact spectroscopy (PCS) \cite{Hasselbach1992}, and neutron scattering measurements \cite{WiebeURS,Janik2009}, as well as in other thermodynamic probes (heat capacity, thermal expansion and ultrasound velocity), which are sensitive to changes in the elastic constants of the crystal lattice, also suggest this temperature scale \cite{Schlabitz,vanDijkURS1997,deVisserURS1986,wolfJLTP1994}. Ultrafast and conventional optical spectroscopy have also found a suppression of low energy spectral weight below 30 K \cite{BonnURL1988,LiuPRB2011,OpticalHybridGapURS2011}.

Under hydrostatic pressure, $T_{HO}$ increases, and above 4.7 kbar \urusi\ undergoes a phase transition to a large moment antiferromagnetic (LMAF) ground state \cite{KoharaURSinhomogeneity,HassingerURSpressure}.   Little is known, however, about the behavior of $T_0$ under pressure.  In order to investigate how the suppression of the density of states evolves, we performed $^{29}$Si NMR experiments  up to 9.1 kbar in the paramagnetic region of the phase diagram.  We find that this suppression persists with the application of pressure, and that $T_0$ increases slightly.

\section{Sample Preparation and Pressurization}


$^{29}$Si NMR ($I=1/2$, 4.7\% natural abundance) measurements were conducted under pressure on a single crystal at 2.0 kbar and on an aligned powder at 9.1 kbar. Details of the sample preparation of both the single crystal and aligned powder are given in Ref. \cite{ShirerURSPRB2012}. The signal to noise ratio in a single-crystal in a pressure cell is poor because the high conductivity of the sample screens out the radiofrequency fields of the pulsed NMR experiment, and small crystals are necessary due to the confined space within the pressure cell. As shown previously, the \slrrtext\ measured in an aligned powder is identical to that in a single crystal, and the data is reproducible across different samples. Therefore we switched to the aligned powder for the higher pressure experiments in order to enhance signal to noise. Pressure was applied using a Cu-Be piston-cylinder clamp cell, with daphne oil (Idemitsu Co., Ltd., type 7373) as a pressure medium.

\begin{figure}
\includegraphics[width=\linewidth]{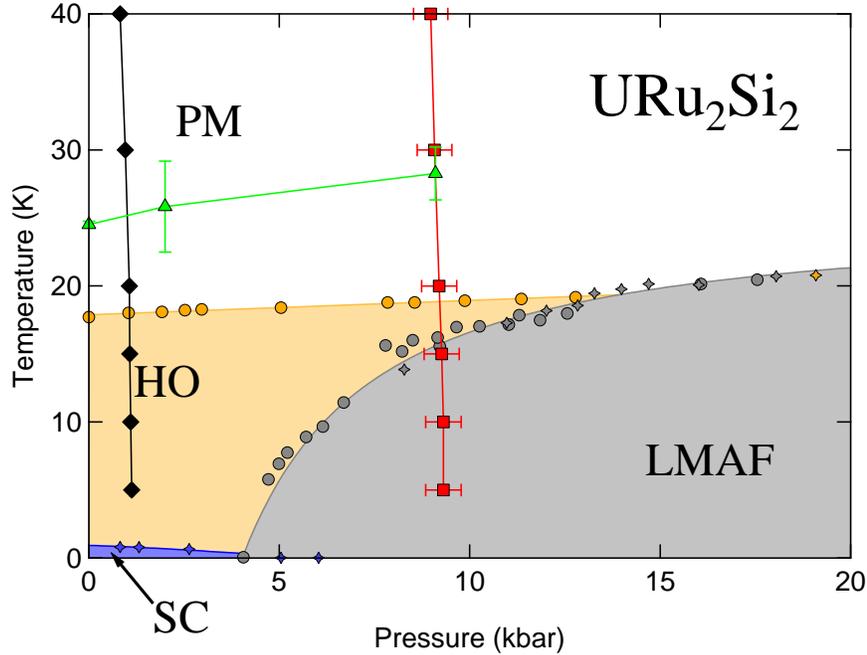}
\caption{(Color Online)
The phase diagram of \urusi, reproduced from \cite{HassingerURSpressure}.  The temperature dependence of the pressure as measured by ruby fluorescence is shown for applied pressures of 2.0 kbar (black diamonds) and 9.1 kbar (red squares). Note that in the temperature range of interest, the pressure varies by less than 0.1 kbar.  The fit parameter $\Delta$ extracted from the fits to the \slrr\ data (green triangles) is shown as a function of pressure.
}
\label{fig:phasedia}
\end{figure}

In order to accurately measure the pressure, we first tested the cell with two different types of manometers:  ruby fluorescence at 532 nm and $^{63}$Cu NQR of Cu$_2$O. The fluorescence was measured by incorporating an optical fiber (200 $\mu$m diameter with Al cladding) in the feedthrough to the pressure cell. A ruby chip of diameter $<300 \mu$m was fixed to the end of the fiber with epoxy. 
The temperature dependence of the pressure in the cell measured using both the shift of the ruby fluorescence line, and the shift of the $^{63}$Cu NQR frequency in a single crystal of Cu$_2$O for various load pressures applied to the cell at room temperature \cite{syassenRubyPress2008, reyesCuNQR1992,KitagawaCellJPSJ2010}.
Both measurement techniques indicate that the pressure changes with temperature on the order of $\sim 10\%$ between room temperature and 4K. This change is likely due to differences in thermal contraction between different elements of the pressure cell. Curiously, the NQR measurement consistently gives a slightly lower pressure ($\sim 10-15\%)$ than the ruby fluorescence, albeit with a similar temperature trend.  We attribute this discrepancy to differences in calibrations.  For example, the NQR calibration is based on low temperature measurements of the $T_c$ of superconducting Sn, whereas the ruby fluorescence calibration is based on x-ray scattering at room temperatures.  The difference is not due to inhomogeneous pressure within the pressure cell, as
we found no broadening of either the ruby fluorescence spectrum or the NQR spectrum up to 9.1 kbar \cite{syassenRubyPress2008}.   Based on these results, we decided to use the ruby fluorescence for the \urusi\ NMR study.  This approach to measuring the pressure is superior because it provides information about the pressure as a function of temperature and does not require that Cu$_2$O be included within the NMR solenoid. 
The pressure measured in a clamp-cell can vary as a function of temperature, particularly at low pressures ($< 10$ kbar). Since \urusi\ (Fig. \ref{fig:phasedia}) exhibits multiple phases in this pressure regime, it is crucial to know the actual temperature dependence of the pressure \emph{in situ} during the experiment. It is important to note that while the freezing of the pressure fluid  around 200K causes a decrease in the applied pressure before it increases below 100K, the pressure in the region of interest for this experiment varies by less than 0.1 kbar.

\section{NMR Results}

\begin{figure}
\includegraphics[width=\linewidth]{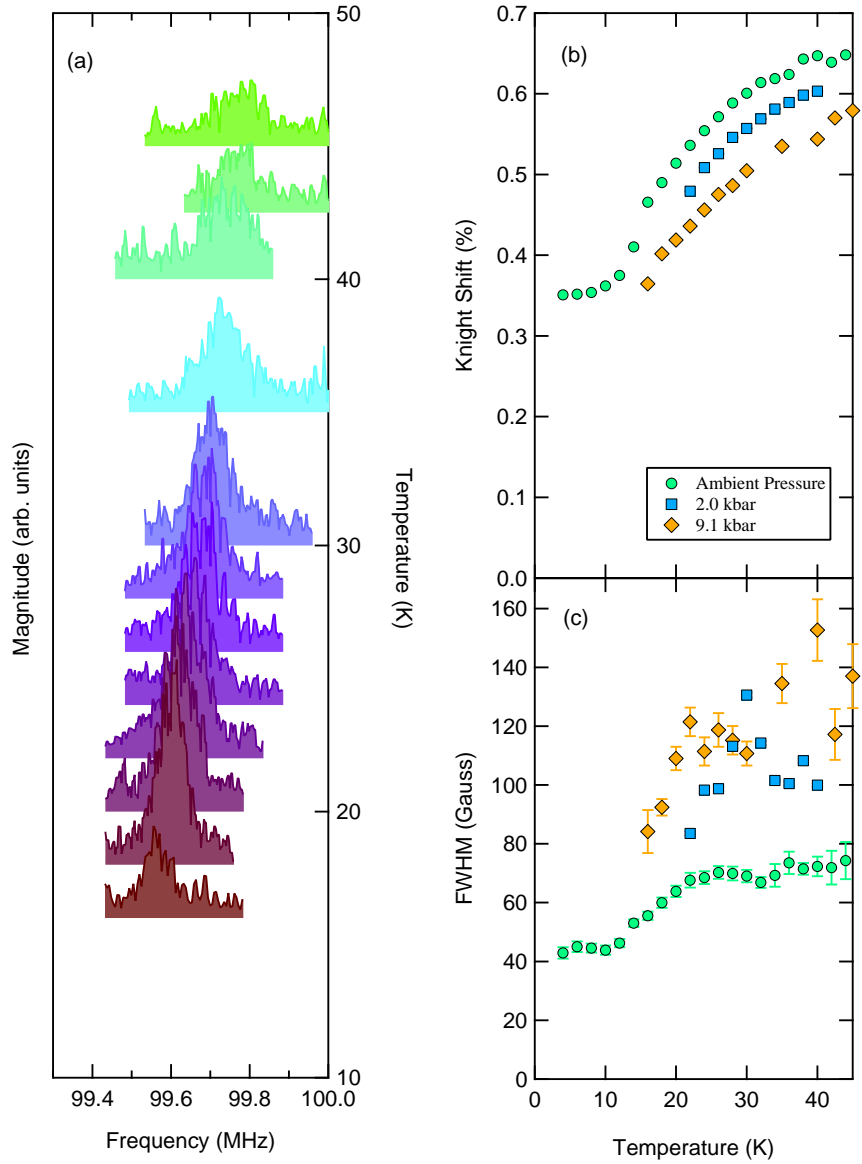}
\caption{(Color Online) (a) Temperature dependent spectra acquired at 9.1 kbar applied pressure. (b) Knight shift of the ambient pressure (green circles), 2.0 kbar (blue squares), and 9.1 kbar applied pressure (orange diamonds). 
The data at ambient and at 2.0kbar was acquired in the single crystal, and the 9.1kbar data was acquired in the aligned powder.
}
\label{fig:KS_FWHM}
\end{figure}

\paragraph{Spectra and Knight Shift} $^{29}$Si NMR spectra were acquired by sweeping frequency at $H_0 = 11.72$ T in  a high homogeneity NMR magnet with the field  parallel to the c--axis of the crystal. In this field, $T_{HO}$ is suppressed to 16 K \cite{JaimeURSPRL2002}.  The $^{29}$Si spectra were measured by spin echoes, and the signal-to-noise ratio was enhanced by summing several (16 to 24) echoes acquired via a Carr-Purcell-Meiboom-Gill (CPMG) pulse sequence with $t_{90} = 4.5\mu$s and an acquisition time of $108\mu$s. The CPMG sequence requires that the spin-spin relaxation time, $T_2 = 7.692$ms for \urusi in the normal state, is long compared to the acquisition of pulses, as is the case here. At ambient pressure, the spin-lattice-relaxation rate $T_1^{-1}$ was measured as a function of the angle between the alignment axis and the magnetic field $\mathbf{H}_0$ in order to properly align the sample, since $T_1^{-1}$ is a strong function of orientation with a minimum for $\mathbf{H}_0 \parallel c$. Figure \ref{fig:KS_FWHM}(a) shows the temperature dependence of the $^{29}$Si spectrum at 9.1 kbar. A comparison of the Knight shifts at ambient pressure, 2kbar, and  9.1 kbar are plotted in Fig. \ref{fig:KS_FWHM}(b). The Knight shift probes the magnetic susceptibility through the hyperfine coupling, and the reduction of the Knight shift with applied pressure is consistent with measurements of the static magnetic susceptibility under pressure \cite{PfleidererPRB2006}.
As previously observed at ambient pressure, the 30K anomaly seen in the \slrrtext\ is not manifest in the Knight shift under pressure \cite{ShirerURSPRB2012,ShirerPNAS2012}.  Fig. \ref{fig:KS_FWHM}(c) shows the temperature dependence of the full width half maximum of the Si resonance. The linewidth increases with applied pressure, and appears to have a peak in the powder sample around $T \sim 25K$. This peak occurs  at a slightly greater temperature in the single crystal, which likely arises from variations between the samples \cite{BernalURS, KambeURS2013}.


\begin{figure}
\includegraphics[width=\linewidth]{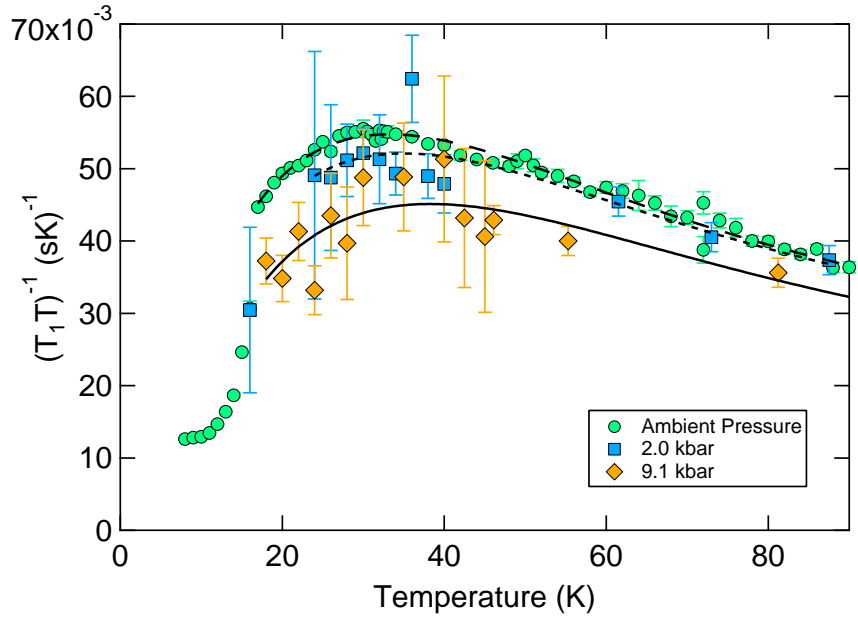}
\caption{(Color Online)} $(T_1T)^{-1}$ versus temperature for the powder and single crystal samples under pressure. At each pressure, \slrr\ has a maximum at $T \sim 30K$, indicating that there is a reduction in the density of states above the hidden order transition that is robust under pressure up to 9.1 kbar.  The data at ambient pressure is reproduced from \cite{ShirerURSPRB2012}. Fits to the data are shown as black lines, and are discussed in the text.
\label{fig:T1Tinv}
\end{figure}

\paragraph{Spin Lattice Relaxation Measurements} The \slrrtext\ was measured by inverting the magnetization and measuring the recovered magnetization by summing the echoes from a CPMG sequence. Fig. \ref{fig:T1Tinv} shows $(T_1T)^{-1}$ versus temperature for several different pressures. The error bars are greater under pressure because of the smaller sample size and reduced signal to noise ratio. Nevertheless, the data indicate a maximum close to 30 K that is characteristic of the reduction of density of states below $T_0$ that persists under pressure. For concreteness, we fit the data in the paramagnetic region to the expression \cite{ShirerURSPRB2012}:
\begin{equation}
\frac{1}{T_1T} = A\left(1-\tanh^2\left(\frac{\Delta}{2 T}\right)\right)\frac{1}{\Delta+T\exp(-4\Delta/T)} + B,
\label{eqn:T1Tinv}
\end{equation}
where $A$, $B$ and $\Delta$ are variable parameters, shown as black lines in Fig. \ref{eqn:T1Tinv}. Here $\left(1-\tanh^2\left({\Delta}/{2 T}\right)\right)$ is a phenomenological form of the normalized joint density of states (JDOS) at the Fermi energy $E_F$, which accounts for any $T$-dependent suppression due to a pseudogap, and $(\Delta+T\exp(-4\Delta/T))^{-1}$ is a crossover function due to 2D antiferromagnetic psuedospin fluctuations \cite{ShirerURSPRB2012, bangT1pumga5}. Although the error bars are significant due to the poor signal to noise, the increasing trend in $\Delta$ as a function of pressure is evident, as shown in Fig. \ref{fig:phasedia}.


\section{Discussion}

The physical significance of $T_0$ remains unclear, but may be related either to the the Kondo hybridization gap or to the onset of fluctuations of the hidden order. In either case, it is reasonable to expect that the energy scale increases with increasing pressure.  Pressure increases the overlap between the U $f$ orbitals and the conduction electrons, leading to an enhanced Kondo scale.  A similar increase was recently observed in NMR experiments in CeRhIn$_5$ under pressure \cite{Lin2015}.
Furthermore, $T_{HO}$ also increases with pressure, so the onset temperature for fluctuations would naturally increase. Curiously, the Knight shift is suppressed by about $19\%$ at 9.1 kbar between 23-30K.  As shown in Ref. \cite{PfleidererPRB2006}, the bulk susceptibility is reduced by about $23\%$ by 15.9 kbar over the same temperature range.  Assuming a linear behavior of the suppression, the bulk susceptibility should be reduced by only $13\%$.  Therefore, the data suggest that the Knight shift has suppressed more than would be expected based on the susceptibility.  A possible explanation for this discrepancy is that the transferred hyperfine coupling between the U moments and the Si nuclei increases under pressure.  Similar pressure-dependent hyperfine couplings were observed in CeRhIn$_5$ \cite{Lin2015}. Further experiments on \urusi\ under pressure are challenging, however, because the spin-spin decoherence rate, $T_2^{-1}$ increases close to the LMAF phase. As a result, fewer CPMG echoes can be summed, and the signal to noise ratio decreases. It is possible that future experiments using isotopically enriched Si may provide more detailed information about the evolution of of the temperature scales, $T^*$ and $T_0$ as a function of pressure in a single crystal.

\section*{Acknowledgements}

Work at UC Davis and LANL was supported by the UC Lab Research Fee Program and the National Nuclear Security Administration under the Stewardship Science Academic Alliances program through DOE Research Grant \#DOE DE-FG52-09NA29464.

\section*{References}

\bibliography{URu2Si2_Pressure_NMR_V2}
\end{document}